\UseRawInputEncoding
\documentclass[twocolumn]{aastex61}
\pdfoutput=1
\usepackage{amsmath,amstext}
\usepackage[T1]{fontenc}
\usepackage{apjfonts} 
\usepackage[figure,figure*]{hypcap}
\usepackage{graphicx}
\usepackage{lineno}
\shorttitle{LMC Abundances}
\shortauthors{R.\, Asa'd, et al.}

\begin{document}

\title{Detailed Chemical Abundances of Star Clusters in the Large Magellanic Cloud}

\author{Randa Asa'd}
\affiliation{American University of Sharjah, Physics Department, P.O. Box 26666, Sharjah, UAE}

\author{S. Hernandez}
\affiliation{AURA for ESA, Space Telescope Science Institute, 3700 San Martin Drive, Baltimore, MD 21218, USA}

\author{A. As'ad}
\affiliation{King Abdullah School for Information Technology, University of Jordan, Amman, Jordan}

\author{M. Molero}
\affiliation{Dipartimento di Fisica, Sezione di Astronomia, Universit\`{a} degli studi di Trieste, Via G.B. Tiepolo 11, I-34143 Trieste, Italy}

\author{F. Matteucci}
\affiliation{Dipartimento di Fisica, Sezione di Astronomia, Universit\`{a} degli studi di Trieste, Via G.B. Tiepolo 11, I-34143 Trieste, Italy; INAF, Osservatorio Astronomico di Trieste, Via Tiepolo 11, I-34131 Trieste, Italy; INFN, Sezione di Trieste, Via Valerio 2, I.34127 Trieste, Italy}

\author{S. Larsen}
\affiliation{Department of Astrophysics/IMAPP, Radboud University, P.O. Box 9010, 6500 GL Nijmegen, The Netherlands}

\author{Igor V. Chilingarian}
\affiliation{Smithsonian Astrophysical Observatory, 60 Garden St. MS09, Cambridge, MA, 02138, USA}
\affiliation{Sternberg Astronomical Institute, M.V. Lomonosov Moscow State University, 13 Universitetsky prospect, Moscow, 119991, Russia}

\correspondingauthor{Randa Asa'd}
\email{raasad@aus.edu}

\begin{abstract}

We derive the first detailed chemical abundances of three star clusters in the Large Magellanic Cloud (LMC), NGC1831  (436$\pm$22 Myr), NGC1856 (350$\pm$18 Myr) and [SL63]268  (1230$\pm$62 Myr) using integrated-light spectroscopic observations obtained with the Magellan Echelle spectrograph on Magellan Baade telescope. 
We derive [Fe/H], [Mg/Fe], [Ti/Fe], [Ca/Fe], [Ni/Fe], [Mn/Fe], [Cr/Fe] and [Na/Fe] for the three clusters. Overall, our results match the LMC abundances obtained in the literature as well as those predicted by detailed chemical evolution models. 
For clusters NGC1831 and NGC1856, the [Mg/Fe] ratios appear to be slightly depleted compared to [Ca/Fe] and [Ti/Fe]. This could be hinting at the well-known Mg-Al abundance anti-correlation observed in several  Milky Way globular clusters. We note, however, that higher signal-to-noise observations are needed to confirm such a scenario, particularly for NGC 1831. We also find a slightly enhanced integrated-light [Na/Fe] ratio for cluster [SL63]268 compared to those from the LMC field stars, possibly supporting a scenario of intracluster abundance variations. We stress that detailed abundance analysis of individual stars in these LMC clusters is required to confirm the presence or absence of MSPs.\\

\end{abstract}


\section{Introduction and Motivation}

Star clusters are key observational tools for understanding  stellar astrophysics and evolution. Determining the abundances of elements resulting from different nucleosynthetic processes such as Type I supernovae (producing Fe-peak elements, for instance Sc, V, Cr, Mn , Fe, Co and Ni), winds from evolved stars, core collapse supernovae (producing $\alpha$-elements: O, Ne, Mg, Si S, Ar, Ca and Ti), can provide us with information about these complex mechanisms. Detailed chemical abundances of stellar populations are also used to uncover information about the chemical enrichment history of the host galaxy \citep{McWilliam97, Worthey98, Matteucci03}.   \\
In order to reveal the chemical enrichment history of a galaxy, we need accurate ages and abundances of star clusters. Several studies \citep[e.g.][and references therein]{Asad16, asad20, Goudfrooij21, asad21, asad22} discussed the precision of age estimates of star clusters based on integrated spectra. \\
Knowing the abundances of star clusters is also crucial in understanding their own formation and evolution, especially with the discovery of the Multiple Stellar Populations (MSP) phenomenon defined as star-to-star variations in the inferred abundances of light elements \citep[He, C, N, O, Na, Al; e.g.][]{Carretta09}. The discovery of MSPs in stellar clusters continues to challenge our view of star clusters as being simple stellar populations. This phenomenon is one of the most puzzling ones in the field of star clusters and it is still not well understood.

The Large Magellanic Cloud (LMC) is a nearby \citep[51 kpc;][]{Wagner-Kaiser17} irregular galaxy. Due to its proximity, the LMC is an ideal laboratory for studying star clusters in great detail. Although the stellar clusters in the LMC resemble the Galactic globular clusters (GCs) in both shape and population, these extragalactic clusters have ages similar to the open clusters of our galaxy \citep{Bergh91}.
Old globular clusters have proven to be useful tools for tracing both the early formation and the star formation history of the LMC \citep{Olszewski91, Geisler+(1997b), Hill00}. 
These studies and others show that metallicity and age distributions of the cluster population in the LMC are indeed bimodal, with a well-defined gap between 3-4 Gyr and 10-12 Gyr ago. This bimodality has been interpreted as the signature of two main bursts of star formation, an early one producing the 12-15 Gyr old clusters and a more recent one taking place 3 Gyr ago. The most recent burst of star formation was possibly triggered by tidal interaction of the LMC with the Milky Way \citep[See discussion in][]{Matteucci12}. However, we note that recently \cite{Gatto20} identified 16 candidate clusters with estimated ages falling in the so-called age gap, located in the outskirts of the LMC.

Additionally, the confirmation of the presence of MSPs in the GCs in this extragalactic environment through both photometric and spectroscopic studies, make the LMC even more intriguing \citep[see e.g. ][and references within]{Milone09, Piatti20}. The LMC is unique compared to the MW because it allows for the exploration of the presence of MSPs not only in ancient GC but also in intermediate-age clusters. A detailed review of these relatively new results is provided in \citet{Bastian18}. \par

In order to get the full picture of the star formation history and chemical evolution of galaxies, we need to combine existing studies of chemical abundances of old stellar population with those of younger populations. At present there are several different tools available to study the present-day chemical state of galaxies, such as measurements of H II regions \citep{Searle71, Rubin94, Stasiska05}, analysis of the integrated light of star clusters \citep{Larsen12, Hernandez17, Hernandez18, Hernandez19, Bastian19, Bastian20} and analysis of evolved massive stars \cite [See for example][]{Davies15, Origlia19, Asad20EO}. In an effort to further investigate the chemical evolution of the LMC and its stellar clusters, we performed a detailed chemical study of three young massive clusters (YMCs) in this nearby irregular galaxy. \par
In this paper we present the analysis of the integrated light of the YMCs NGC1831, NGC1856 and [SL63]268. We measure for the first time detailed abundances of $\alpha$ (Ti, Mg, and Ca), light (Na), and Fe-peak elements (Ni, Mn and Cr) in these three extragalactic clusters. The paper is organized as follows: in Section \ref{sec:obs} we describe the spectroscopic observations and data reduction; in Sections \ref{sec:method} and \ref{sec:models} we detail the integrated-light analysis technique used to infer the chemical abundances, and the stellar atmospheric models used in generating the synthetic integrated-light spectra, respectively; in Sections \ref{sec:results} and \ref{sec:discussion} we present our results and discuss our findings in the context of chemical evolution and MSPs; and lastly in Section \ref{sec:summary} we summarize our conclusions.  

\section{Observations and Data Reduction}\label{sec:obs}

The observations analyzed as part of this work include three LMC clusters in the age range from 300 Myr to 1.5 Gyr, observed with the 6.5 m Magellan Baade telescope in 2016 November, using the intermediate resolution Magellan Echelle (MagE, Marshall et al. 2008) spectrograph. We used for our observation the 0.5 arcsec wide slit; with resolution (R) =7000;  and wavelength range: 3300 $< \lambda$(\AA) $<$ 9500. 
We used the scan mode with 2$\times$900sec exposures for NGC1831,  2$\times$900sec exposures for NGC1856 and 2$\times$1800sec exposures for [SL63]268. The signal-to-noise S/N ratio for the redder side of [SL63]268 monotonically increases from about 40~\AA$^{-1}$ to 80~\AA$^{-1}$ at wavelength between 4000~\AA\ and 6700~\AA, while for the other two clusters it stays nearly constant over wavelength at 60~\AA$^{-1}$ for NGC~1831 and 160~\AA$^{-1}$ for NGC~1856.

We  refer the interested reader to \cite{Chilingarian18} for more details about the observations.  
In order to collect an integrated spectrum, we scanned a cluster across the slit using the nonsideral tracking. 
The data reduction was done by developing a dedicated data reduction pipeline that merges Echelle orders and produces a sky-subtracted flux-calibrated two-dimensional spectrum corrected for telluric absorption and its corresponding flux uncertainty frame. The spectrum is then collapsed along the slit to obtain a one-dimensional flux-calibrated data product in the wavelength range 3700$ < \lambda\:$ (\AA) $<6800$.

\section{Method}\label{sec:method}

We use the software developed for GCs by \citet{Larsen12}, and later applied to YMCs by \citet{Hernandez17}, to obtain detailed abundances for our LMC sample. This technique relies on existing age estimates for the stellar populations under study, to select the best initial isochrone.\par
We generate theoretical PARSEC isochrones \citep{Bressan12} adopting the ages estimated for our sample by \citet{Chilingarian18}, and an initial metallicity of [m$/$H] = -0.4. 
We adopt a \cite{Salpeter55} initial mass function (IMF) and a lower mass limit of 0.4 M$_{\odot}$ when extracting the stellar atmospheric parameters (e.g., effective temperature, surface gravity, mass) from the isochrones. Lastly, we account for the microturbulent velocity component, $v_t$, assigning different values depending on the effective temperature ($T_{\rm eff}$) of the individual stars: $v_t$ = 2 km s$^{-1}$ for stars with $T_{\rm eff} <$ 6000 K, $v_t$ = 4 km s$^{-1}$ for stars with 6000 $< T_{\rm eff} <$ 22 000 K, and $v_t$ = 8 km s$^{-1}$ for stars with $T_{\rm eff} >$ 22 000 K \citep{Hernandez17, Hernandez18}.\par 
We note that the analysis of integrated-light observations of stellar clusters is subject to stochastic fluctuations in the number of stars of a given stellar type falling in the scanned area. The effects of random IMF sampling in the type of analysis described here have been assessed in previous studies \citep[e.g.,][]{Larsen12, Larsen17}. \cite{Larsen17} has tested the uncertainties introduced by the stochastic sampling of the HRD. They produced Monte-Carlo simulations where the same number of stars as those present within the scanned areas were sampled at random from the full CMDs, and performed a similar abundance analysis on these randomly sampled CMDs. Overall, the stochastic sampling experiment in \cite{Larsen17} indicated that the stochastically induced (1-$sigma$) uncertainty on overall metallicities is $<$ 0.1 dex. \par
Using the software by \citet{Larsen12} we first obtain an estimate of the radial velocity of the individual star cluster, by fitting small wavelength windows of 200\AA $ $ from 4000\AA $ $ to 6200\AA. This wavelength range was selected as it provided a reasonable number of bins and excluded the lowest S/N regions at wavelengths $\lesssim$4000 \AA. We estimate the mean radial velocities and adopt these values for the rest of our abundance analysis (see the first row in Table \ref{All_results}). Our estimated values are comparable to those of \cite{Chilingarian18}, who obtained radial velocities of 278.8$\pm$0.6 km/s, 268.6$\pm$0.2 km/s and 266.0$\pm$0.5 km/s for NGC1831, NGC1856 and [SL63]268, respectively. \par
To match the resolution between the synthetic integrated-light models and that of the spectroscopic observations, we fit for the best Gaussian dispersion, $\sigma_{\rm sm}$. 
The scaling of the synthetic spectrum is determined by fitting the ratio of the synthetic and observed spectra with a spline or polynomial functions, depending on the size of the wavelength bin. In Section \ref{sec:continuum}, we present a brief discussion on the sensitivity of our analysis to the scaling procedure.\par

In this work, [Z] is defined as a scaling parameter relative to solar composition and applied to all of the specified abundances. This means that [Z] is a measure of the integrated abundances of the different chemical elements. In the current work we have adopted the solar composition from \citet{Grevesse98}.\par
We begin the abundance analysis by simultaneously fitting for the best $\sigma_{\rm sm}$ and [Z]. We fit each spectrum using 200-\AA\: bins scanning the wavelength range between 4000 \AA\: and 5800 \AA\:. This wavelength region provided sufficient information for deriving the initial scaling parameter, [Z], and at the same time allowed us to exclude the lower S/N regions at $\lesssim$4000 \r{A}.
A direct comparison is made between the model spectrum and the science observations by performing a $\chi^2$ minimization. The continua of the model and observed spectra are matched using a cubic spline with three knots. The process is repeated modifying the variables accordingly until the best match is determined.

We then proceed to measuring the abundance of a number of individual elements starting with those having the highest number of lines, while keeping the overall metallicity and the smoothing parameters fixed.

We first measure Fe then Ti and Ca, followed by the rest of the elements.  We masked the bad pixels around 6300 \AA\: for NGC1831, NGC1856 and [SL63]268. We used the optimised wavelength windows tailored for each element introduced by \citet{Hernandez17} in order to ameliorate the effects of strong blending.\\
Similar to the procedure followed for the overall metallicity estimation, we use a cubic spline with three knots to match the continua of the model and observed spectra for windows $\geq$ 100 \AA\:. For bins narrower than 100 \AA\: we use a first-order polynomial instead.\\

\subsection{Scaling of the continuum level}\label{sec:continuum}
 \citet{Larsen12, Larsen14} discuss extensively the proper scaling of the continuum level. To avoid problematic continuum placements due to weak features present in the data, they have identified continuum regions in the observed spectrum of Arcturus, and use only these regions free of absorption features when scaling the model and science observations. Similar to the analysis of intermediate-resolution observations of YMCs in \citet{Hernandez17} we made use of these pre-defined continuum regions when scaling the model and observations.\par
  To get a better sense of the sensitivity of our analysis to the continuum scaling procedure, we estimated the overall metallicities for NGC 1831, the cluster with the lowest S/N values in our sample, applying three different methods: (1) our standard method of using the predefined continuum regions from \citet{Larsen12, Larsen14} and \citet{Hernandez17}, and applying a spline function with three knots to determine the scaling of the model spectrum, (2) using the predefined continuum regions, along with a first order polynomial to determine the scaling of the model spectrum, (3) applying no restrictions on the continuum pixels and using a spline function with three knots to determine the scaling of the model spectrum. We found differences of 0.05 dex in the overall metallicities inferred using method (2), compared to the value obtained with our standard approach (method 1). Similarly, we found differences of 0.06 dex when applying method (3), in comparison to our adopted approach. Given that the overall S/N values for the other two clusters are higher, these differences provide an upper limit to the uncertainties introduced by the continuum scaling procedure.

\section{Models}\label{sec:models}
The \citet{Larsen11} technique produces synthetic integrated-light spectra by co-adding individual spectra  generated for each of the stars in the cluster.
The synthetic spectra are generated by creating a series of high spectral resolution models that include all evolutionary stages present in the star cluster. The software first computes atmospheric models using ATLAS9 \citep{Kurucz70} for stars with T$_{\rm eff} > $ 3500 K and using MARCS \citep{Gustafsson08}
for stars with T$_{\rm eff} < $ 3500 K. \\ 
The atmospheric models are then used to create synthetic spectra with SYNTHE \citep{Kurucz79, Kurucz81} for the ATLAS9 models and TURBOSPECTRUM \citep{Plez12} for MARCS models. The model spectra are created at high resolution of R = 500,000 and then degraded to match the observations. For the synthetic spectral computations, we use the line lists by \cite{Castelli04} which include hyperfine splitting for a few Mn I lines. 

\begin{figure*}
\centering
\makebox[0pt]{\includegraphics[scale=0.55]{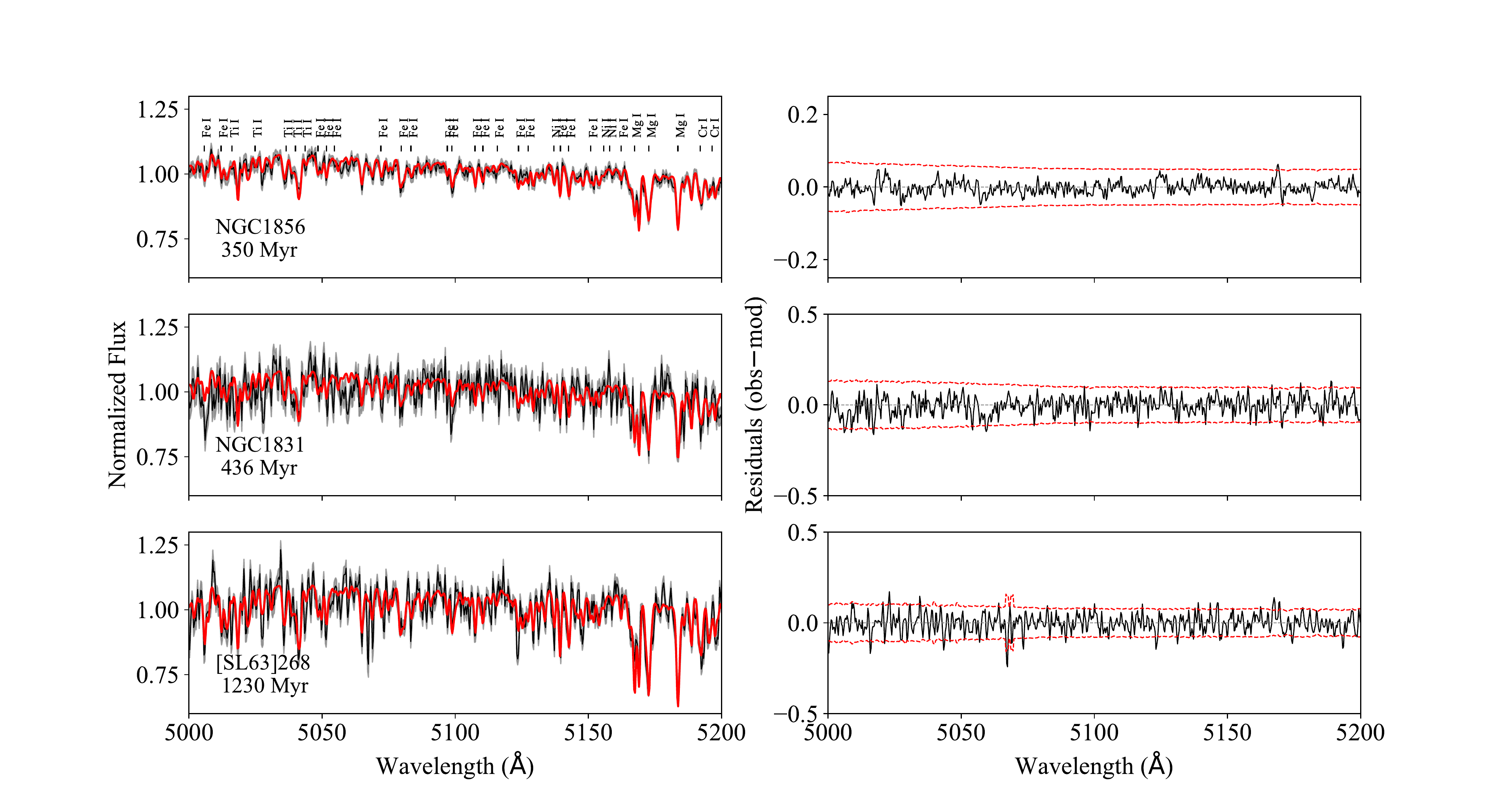}}
\caption{Left: Synthesis fits for our sample of YMCs in the LMC. In black we show the MagE spectroscopic observations, along with their uncertainties in grey. In red we show the model fit. We indicate the cluster name and the assumed age below the corresponding spectra. Right: We display the residuals in black. The red dashed lines indicate the 3$\sigma$ error spectrum.}
\label{Fits}
\end{figure*}

\section{Results}\label{sec:results}

In Figure \ref{Fits} we show our synthesis fits for all three clusters, along with their residuals.
The chemical abundances obtained are included in Tables  \ref{Table_NGC1831}, \ref{Table_NGC1856} and \ref{Table_SL268} where we list the wavelength bins used in the abundance measurement of each element, the best fit abundance and their corresponding uncertainties calculated from the $\chi^2$ fit. We used five bins for Fe, four bins for Ca, three bins for Ti, four bins for Cr, five bins for Ni, two bins for Na  and one bin for each of Mg and Mn. The tables show that for several bins the software was unable to converge to a final abundance. 
Table \ref{All_results} presents the weighted average abundances of our clusters and their uncertainties obtained by dividing the standard deviation by $\sqrt{N-1}$, where N is the the number of wavelength bins used for obtaining each abundance. Here $[$Fe/H$]=$ Fe + [Z], where Fe is the mean iron abundance estimated for each wavelength bin listed in Tables \ref{Table_NGC1831}, \ref{Table_NGC1856} and \ref{Table_SL268}.
The other listed elements [X/Fe] were calculated using the equation [X/Fe] = [X/H] - [Fe/H], where [X/H] = X + [Z], and X is the weighted average abundance of element X estimated for each wavelength bin listen in Tables \ref{Table_NGC1831}, \ref{Table_NGC1856} and \ref{Table_SL268}.  \\

\begin{table}
\caption{Our Results}
\label{All_results}
\begin{tabular}{ccccc}
\hline
  & NGC1831 & NGC1856 & [SL63]268 \\ 
\hline

RV (km s$^{-1}$) & 290$\pm$7 & 279$\pm$4 & 278$\pm$4 \\ 
$\sigma_{\rm sm}$ (km s$^{-1}$) & 23.9 & 26.9 & 24.6 & \\
$[$Z$]$ & -0.418$\pm$0.07 & -0.574$\pm$0.06 & -0.51$\pm$0.04 & \\
$[$Fe/H$]$ & -0.375$\pm$0.12 & -0.455$\pm$0.11 & -0.506$\pm$0.1 &  \\
$[$Ca/Fe$]$ & 0.814$\pm$0.41 & 0.375$\pm$0.2 & -0.277$\pm$0.39 & \\
$[$Na/Fe$]$ & 0.023$\pm$0.77 & 0.093$\pm$0.38 & 0.357$\pm$0.08 & \\
$[$Mg/Fe$]$ & 0.082$\pm$0.16 & -0.074$\pm$0.13 & 0.07$\pm$0.11 & \\
$[$Ti/Fe$]$ & 0.547$\pm$0.42 & 0.262$\pm$0.2 & 0.233$\pm$0.19 & \\
$[$Cr/Fe$]$ & 0.002$\pm$0.5 & 0.132$\pm$0.16 & 0.111$\pm$0.0 & \\
$[$Mn/Fe$]$ & 0.323$\pm$0.38 & -0.263$\pm$0.32 & -0.199$\pm$0.21 & \\
$[$Ni/Fe$]$ & 0.318$\pm$0.5 & -0.194$\pm$0.08 & -0.224$\pm$0.41 & \\
\hline
\end{tabular}
\end{table}

\section{Discussion}\label{sec:discussion}

\subsection{Comparing our Results}

In this section, we compare our results with those obtained from previous studies. 
In Table \ref{T3} we list our results for %[Z],
[Fe/H] and [Mg/Fe], as well as those obtained by \citet{Chilingarian18}. \\
\citet{Chilingarian18} used {\sc nbursts} \citep{Chilingarian07}, a full spectrum fitting tool, to obtain the age, Z and [Mg/Fe]. [Fe/H] is calculated using the equation given in \citet{Chilingarian18}: 
\begin{equation}
[\mathrm{Fe/H}] = [\mathrm{Z/H}] - 0.75[\mathrm{\alpha/Fe}] 
\label{eq1}
\end{equation}
where [$\alpha$/Fe] is the obtained [Mg/Fe] from {\sc nbursts}. Overall, the results match reasonably well, with the exception of [Mg/Fe] for NGC1856, where we measure a slightly more depleted ratio than \citet{Chilingarian18}.

\begin{table*}
\caption{Ages, metallicities and abundances of our sample}
\label{T3}
\begin{center}
\begin{tabular}{cccccccc}
\hline
Object & Age$_{\rm Lit}^1$ & [Fe/H]$_{\rm Lit}^2$ &  [Fe/H]&  [Mg/Fe]$_{\rm Lit}^3$ &  [Mg/Fe] \\
\hline

\object{NGC1831} & 436$\pm$22 & -0.39$\pm$0.05 & -0.38$\pm$0.12 & 0.09$\pm$0.06 & 0.08$\pm$0.16 \\
\object{NGC1856} & 350$\pm$18 & -0.47$\pm$0.02 & -0.46$\pm$0.11 & 0.18$\pm$0.03 & -0.07$\pm$0.13 \\
\object{[SL63]268} & 1230$\pm$62 & -0.58$\pm$0.03 & -0.51$\pm$0.1 & 0.09$\pm$0.02 & 0.07$\pm$0.11 \\
\hline \\
\end{tabular}
\end{center}
1. The ages in Myr from \cite{Chilingarian18} are the ones obtained when counting for alpha elements. 
2. [Fe/H] were calculated using Z and [alpha/Fe] from table 2 in \cite{Chilingarian18} and corrected equation \ref{eq1}.
3. [Mg/Fe] are the values of [alpha/Fe] listed in table 2 in \cite{Chilingarian18} as these values were calculated based on Mg only. [Fe/H] and [Mg/Fe] are measured in dex. 

\end{table*}
\label{T3}

\begin{figure*}
\epsscale{1.2}
\plotone{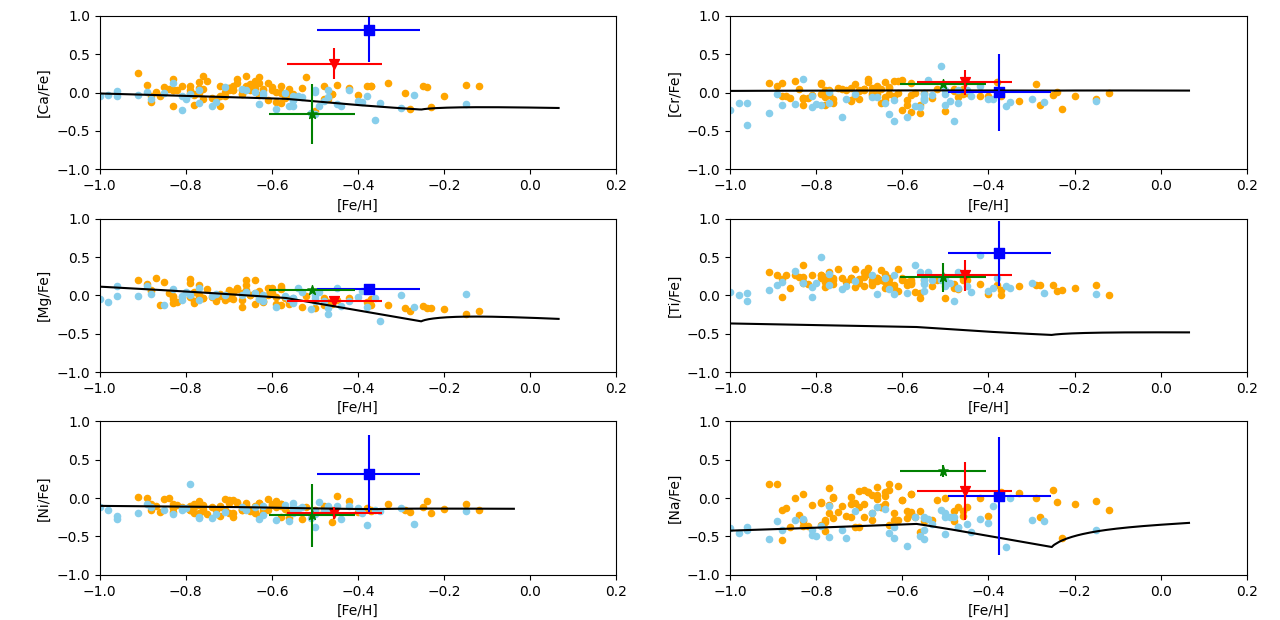}
\caption{Abundance ratios as a function of [Fe/H]. The abundance ratios inferred as part of this work are shown as diamonds (navy square for NGC1831, red triangle for NGC1856 and green star for [SL63]268). The orange and light blue dots are the literature data from field stars in the LMC bar and inner disc abundances, respectively, presented by \citet{VanderSwaelmen13}. We show with a solid black line the chemical evolution models for the LMC.}
\label{Fig1}
\end{figure*}

In Figure \ref{Fig1}, we show our inferred YMC abundances along with those from field stars in the LMC and theoretical models. The abundance ratios inferred as part of this work are shown as different symbols (navy for NGC1831, red for NGC1856 and green for [SL63]268). The orange and light blue dots are the literature data from field stars in the LMC bar and inner disc abundances, respectively, presented by \citet{VanderSwaelmen13} and the black line is the theoretical LMC model. We note that the model was originally scaled to the solar abundances of \citet{Asplund09}, however, we homogenize the model and measured abundances to the single abundance scale of \cite{Grevesse98}. The chemical evolution model adopted for the LMC is similar to that of \citet{Calura03}. The evolution of the gas abundances of several chemical  species is computed in detail. The chemical enrichment from supernovae of all types (II, Ia, Ib, Ic) as well as from AGB stars is considered together with stellar lifetimes. 

For the history of star formation (SF) of the LMC we assume two main bursts, the first between 0 and 5 Gyr and the second at 12 Gyr. The efficiency of SF is 0.1 Gyr$^{-1}$ during the bursts and 0.001 Gyr$^{-1}$ in between. We adopt the duration and number of bursts as suggested by \citet{Harris09} and fix the star formation efficiency to reproduce the abundance pattern, the present time mass of gas and stars as well as supernovae rates.

The star formation rate is defined as :
\begin{equation}
\psi(t)= \nu M_{gas}
\end{equation}

where $\nu$ is the efficiency of star formation. 
We assume that the LMC assembled by infall of primordial gas with the following law:

\begin{equation}
\dot M_{gas,infall}=a X_ie^{-t/\tau}
\end{equation}

where $a$ is a parameter tuned to reproduce the total present time mass of LMC ($M_{LMC}=6$x$10^6$),
$X_i$ is the abundance of the element $i$ in the infalling gas and $\tau=0.5$ Gyr is the infall timescale.

The model considers also gas outflow from the galaxy when the thermal energy produced by SNe and stellar winds equates the binding energy of gas (see \citealt{Bradamante98} for a detailed discussion).

The wind rate is expressed as being proportional to the star formation rate:
\begin{equation}
\dot M_{outflow}= - \omega \psi(t)
\end{equation}
where $\omega=0.25$ is the mass loading factor and is adimensional.

For the stellar yields we adopt those from \citet{Karakas10} for the chemical enrichment from low and intermediate mass stars ($0.8 \le M/M_\odot \le 8$) and from \citet{Kobayashi06} for massive stars ($M>8M_{\odot}$). For Type Ia SNe, here assumed to originate from white dwarfs in binary systems, we adopt the yields from \citet{Iwamoto99}.

The elements considered in this paper are mainly formed in core-collapse SNe (II, Ib, Ic) and Type Ia SNe, so the contribution from AGBs is negligible. It is worth noting that the yields of some elements, such as Ti and Cr, are still very uncertain, as shown by previous works (e.g. \citealt{Romano10, Matteucci21}) and to reproduce the data of the solar vicinity one needs to change them arbitrarily. The reason for such uncertainties can be found in uncertain nuclear reaction rates and treatment of convection in stellar models.

\begin{figure}
\epsscale{1.1}
\plotone{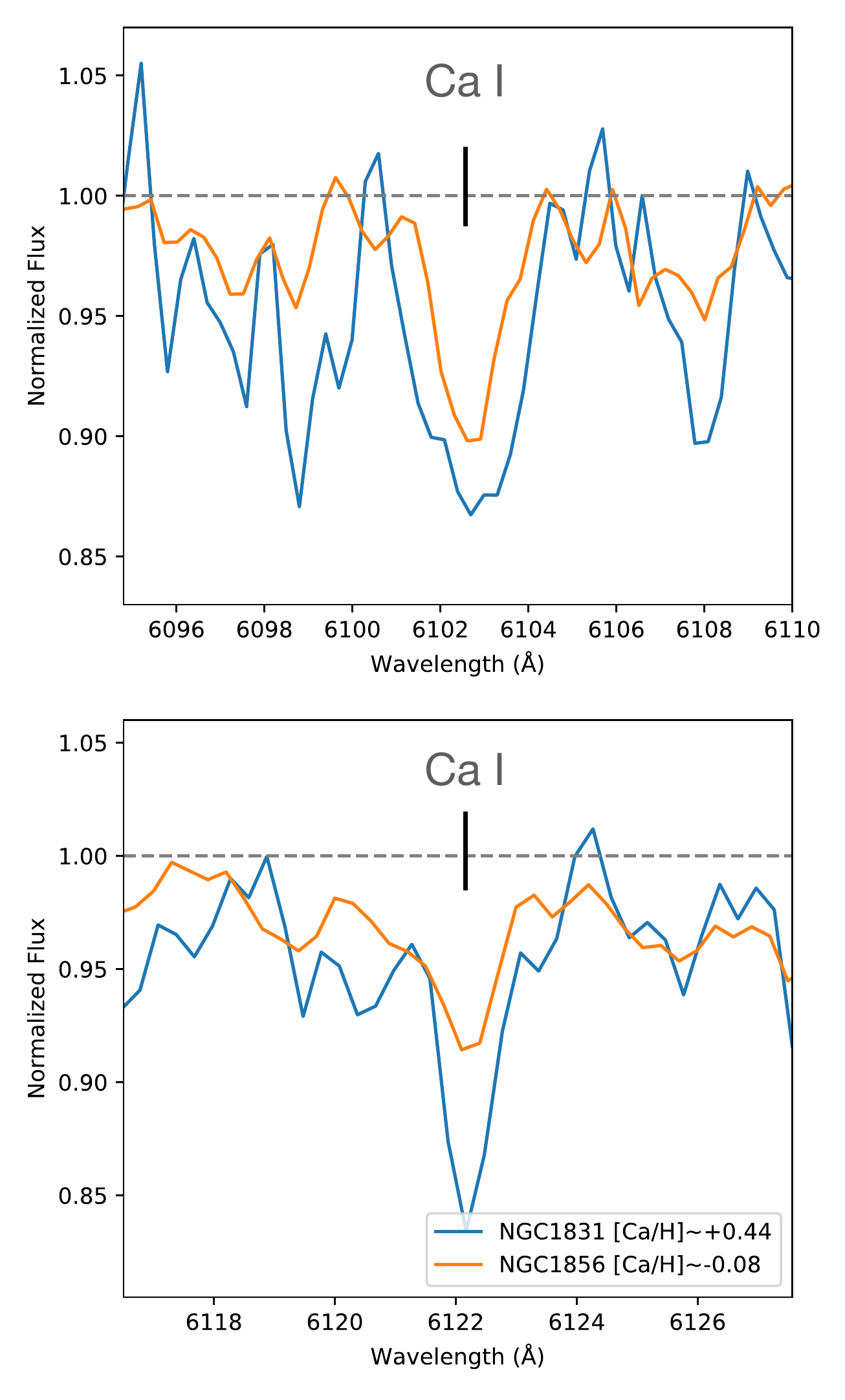}
\caption{Ca spectral lines for NGC1831 (blue) and NGC1856 (orange).}
\label{Fig_Ca}
\end{figure}

Overall, our abundance ratios agree with those predicted by the theoretical model for the available elements. Currently, there are no theoretical models available for Mn, which prevents us from comparing our abundances against predicted values.
The agreement between our inferred abundances and those predicted by the chemical evolution model supports a scenario where the star formation history of the LMC would involve two main bursts. Additionally, the work presented here shows that integrated spectra of star clusters are reliable tools for studying the chemical enrichment history for distant galaxies (where resolved data cannot be obtained). \\
A comparison between our inferred abundances and those measured by \citet{VanderSwaelmen13} for the field stars in the LMC shows that the ratios for Ca, Ni and Ti for NGC1831 are on the high side of the envelope of observed abundances in field. For these three elements we have compared the adopted oscillator strength values (log \textit{gf}) from the study by \citet{VanderSwaelmen13} and those used in our analysis. We find no differences between the log \textit{gf} values for the Ti and Ni lines. For Ca we identify differences of the order of <0.1 dex. Overall, these differences, or the absence of such, most likely indicate that the abundance differences between those from NGC1831 and the field stars are not due to the adopted log \textit{gf} values.\\
We note that from our sample of YMCs, the observations of NGC1831 showed the lowest S/N. This is also reflected in the higher uncertainties for the abundances of this YMC (for example $\sim$0.4 dex for [Ca/Fe]), compared to those for NGC1856 and [SL63]268. 
To confirm if the enhancement in Ca is intrinsic to the YMC NGC1831 in Figure \ref{Fig_Ca} we overplot the Ca lines for NGC1831 and compare them with those for NGC1856. We chose to compare the spectroscopic observations of these two clusters as they have similar ages and metallicities. Figure \ref{Fig_Ca} not only shows that NGC1831 exhibits stronger Ca lines than NGC1856, but it also contrasts the differences in S/N values between the two different datasets. 

\subsection{Multiple Populations}

It has been firmly established that Galactic GCs host MSPs inferred through star-to-star variations in the abundances of some light elements (e.g., He, C, N, O, Na, Al). However, in spite of many observational and theoretical studies, the physical origin of MSPs is still debated. Several scenarios have been proposed to explain this phenomenon, with most implying multiple epochs of star-formation within the cluster; however, none of these scenarios have fully succeeded to reproduce the increasing number of observations obtained in the past decade. Additionally, if YMCs are the young analogs of GCs, we expect to find MSPs in these young stellar populations. We note, however, that MSPs have not been detected in YMCs younger than 2 Gyr \citep{Bastian18, Martocchia21}.\par 
Studies have found intra-cluster Mg variations in several GCs in the MW \citep[see review by][]{Gratton04, Bastian18}. It is possible that these same intra-cluster variations in Mg might be detected in integrated-light studies of GC as (1) lower [Mg/Fe] ratios when compared to those from field stars \citep{Larsen14}, or (2) lower [Mg/Fe] compared to other [$\alpha$/Fe] ratios for the same cluster \citep{Colucci09,Larsen12}. 
In the context of item (1), the middle left panel show that the distribution of [Mg/Fe] is compatible with the Mg abundances of field stars , however, in the context of item (2), our results in Table \ref{All_results} show lower [Mg/Fe] ratios compared to those from [Ca/Fe] and [Ti/Fe] for YMCs NGC1831 and NGC1856 which might be an indication of the presence of MSPs. The upper limit of the  [Mg/Fe] ratio for  NGC1831 when errors are taken into account is 0.242 dex, which is still less than the lower limit of [Ca/Fe] ratio (0.404 dex), but more than the lower limit of [Ti/Fe] ratio (0.127 dex). For NGC1856 however, the upper limit of the  [Mg/Fe] ratio within the error range (0.056 dex) is still significantly less than the lower limit of [Ca/Fe] and [Ti/Fe] ratios (0.175 dex and 0.462 dex, respectively).  \par
\begin{figure*}
\centering
\makebox[0pt]{\includegraphics[scale=0.52]{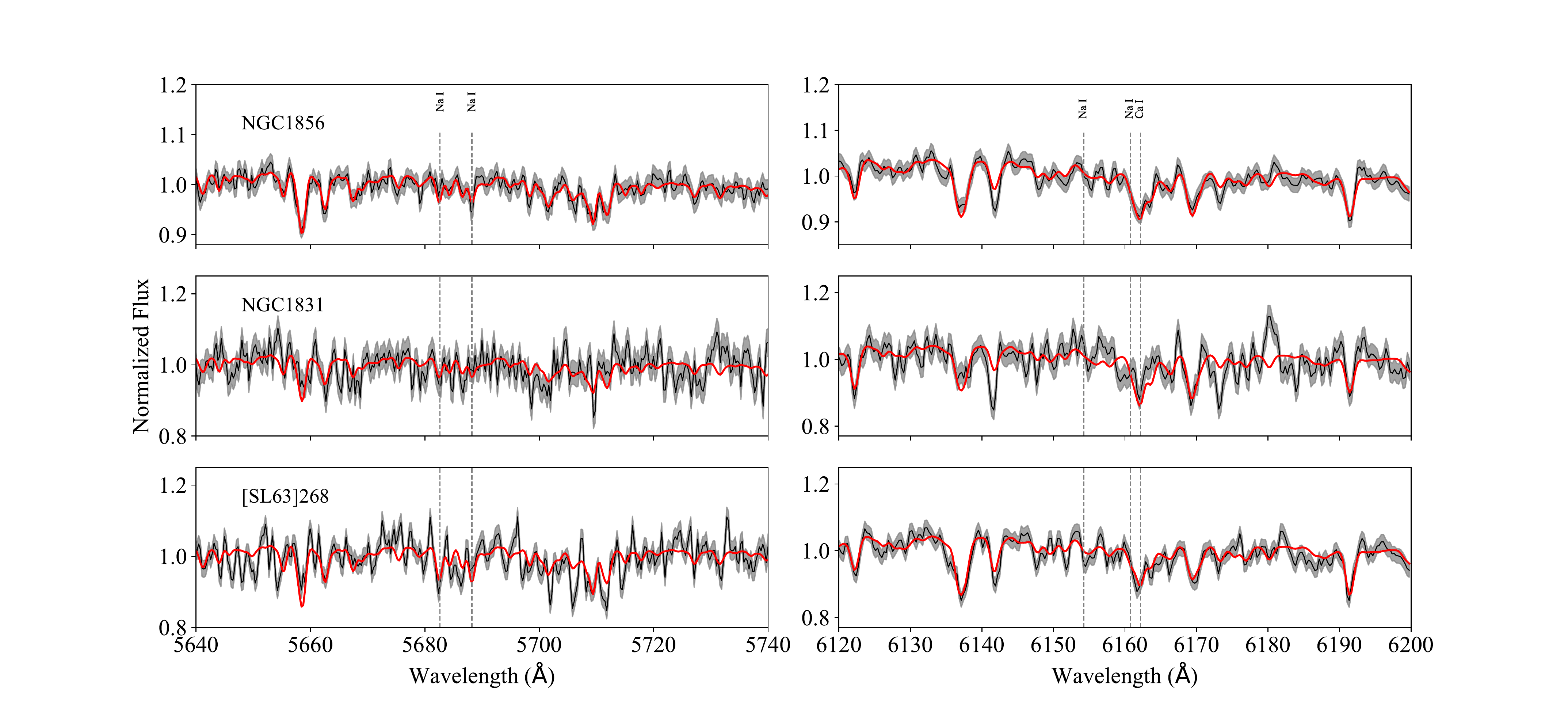}}
\caption{Best synthesis fits for our sample of YMCs. In black we show the MagE spectroscopic observations. In red we display the model fit. We show with dashed vertical lines the location of the Na lines.}
\label{Na}
\end{figure*}

 
In addition to Mg, Na has also exhibited star-to-star variations in Galactic GCs \citep{Gratton04}. These variations in Na abundances are observed as significantly elevated [Na/Fe] ratios in integrated-light studies of GCs when compared to those of field stars \citep{Colucci09, Larsen14, Hernandez18b, Bastian19, Bastian20}. To further investigate the possibility of MSPs in our sample of YMCs, we estimate the Na abundances. For this we analyzed two wavelength bins covering several Na lines. The low S/N observations for NGC 1831 made it very challenging to accurately measure the Na abundances from these two wavelength bins. This is reflected in the large uncertainties of our inferred [Na/Fe] ratio for this particular cluster. In Figure \ref{Na} we display the best fit models and show with dashed vertical lines the location of the Na lines. We highlight that the Na doublet in the 6148-6168 \AA $\:$ wavelength bin is rather weak (log gf $=$ $-$1.228, compared to log gf = $-$0.452 for Na I at 5688 \AA), with one of the lines being blended with a much stronger Ca line.\\
In the lower right panel of Figure \ref{Fig1} we compare our inferred [Na/Fe] ratios for all three clusters, with those from field stars by \citet{VanderSwaelmen13}. 
Overall, this panel shows that the [Na/Fe] ratio is slightly more enhanced in cluster [SL63]268, compared to the abundances of the field stars. The inferred abundance of NGC 1856 appears to be on the upper envelope of observed abundances in the field. The mean value of the [Na/Fe] abundances of the field stars with -0.5 $<$[Fe/H] $<$-0.35 is -0.21, and the inferred values for NGC 1856 and [SL63]268 are 0.093$\pm$0.38 and 0.357$\pm$0.08, respectively. Based on the uncertainties of our measurement for cluster NGC 1856, we are unable to definitively confirm an enhancement in the [Na/Fe] ratio over the abundances of the field stars.\par In the context of MSPs, the Na enhancement in [SL63]268 may be hinting at intracluster abundance variations in a cluster with an age slightly younger than 2 Gyr. [SL63]268 is younger than any LMC/SMC cluster in which evidence of MSPs (in the form of abundance variations, e.g. \citealt{Martocchia21}) has been found so far. We note that in our sample we observe that [SL63]268 shows hints of Na variations but we find no evidence for Mg variations. On the other hand, NGC 1831 shows possible hints of variations in Mg, however, given the large uncertainties in our inferred [Na/Fe] ratio, we are unable to assert the presence or absence of an enhancement of Na over the abundances of the field stars. We highlight that depleted [Mg/Fe] ratios are not observed in all of the clusters in which Na variations are detected \citep{Bastian18}. Locally, only a few Galactic GCs have shown evidence of significantly depleted Mg \citep[e.g.,][]{Mucciarelli12,Carretta14}. More analysis is needed for this sample of LMC clusters to accurately investigate the MSP phenomenon.

\section{Summary}\label{sec:summary}
In this work we perform the first detailed abundance analysis for three LMC YMCs observed with the Magellan Echelle spectrograph on Magellan Baade telescope. We follow the method described in \citet{Hernandez17}. We summarize below the results of our study:\\
(A)  We derive [Fe/H], [Ti/Fe], [Ca/Fe], [Mg/Fe], [Ni/Fe], [Mn/Fe], [Cr/Fe] and [Na/Fe] abundances for NGC1831, NGC1856 and [SL63]268. \\
(B) Our results for [Fe/H] and [Mg/Fe] match reasonably with those derived by \citet{Chilingarian18}.\\
(C) To the best knowledge, this study is the first to derive [Ti/Fe], [Ca/Fe], [Ni/Fe], [Mn/Fe],  [Cr/Fe] and [Na/Fe] abundances for the three clusters of our sample. \\
(D) Overall our results match the LMC abundances obtained in the literature as well as the theoretical model. \\
(E)  The nucleosynthesis of Ti element is very poorly known, further analysis is required to better constrain the theoretical models (See the model line in Figure \ref{Fig1}).  \\
(F) Our results from NGC1831 and NGC1856 show a possible depletion in the [Mg/Fe] abundance compared to [Ca/Fe] and [Ti/Fe].\\
(G) We also observe slightly enhanced [Na/Fe] ratios in cluster [SL63]268, when compared to the abundances of the field stars. This trend can be an indication of intracluster Na variations. \\
Based on observations, current models and theories suggest that the cutting limit between the occurrence and non-accordance of MSPs happens at ages of 2â3 Gyr, indicating that either the cluster age plays a dominant role in the establishment of MSPs or that the phenomenon is present only in low mass stars. The discovery of light element variations in clusters younger than 2 Gyr would provide new insights on the origin of this phenomenon and the parameters that play key roles in it.
Although MSPs in stellar clusters younger than 2 Gyr is an intriguing possibility, we highlight that detailed abundance analysis of individual stars in these LMC clusters is required to confirm such a scenario.

\section*{Acknowledgements}

We are thankful to the referee for the careful review of this manuscript, which helped improve this paper.\\
This work is supported by FRG Grant P.I., R. Asa'd and the Open Access Program from the American University of Sharjah.
RA thanks Radboud University for allowing access to their coma cluster computer facilities that was used in the early stages of this work. 
IC acknowledges the support by the RScF grant 17-72-20119 and by the Interdisciplinary Scientific and Educational School of Moscow University Fundamental and Applied Space Research.
This paper represents the opinions of the authors and does not mean to represent the position or opinions of the American University of Sharjah.

\bibliographystyle{aasjournal}
\bibliography{references}

\appendix
\twocolumngrid
\section{Inferred abundances}
\begin{figure*}
\plotone{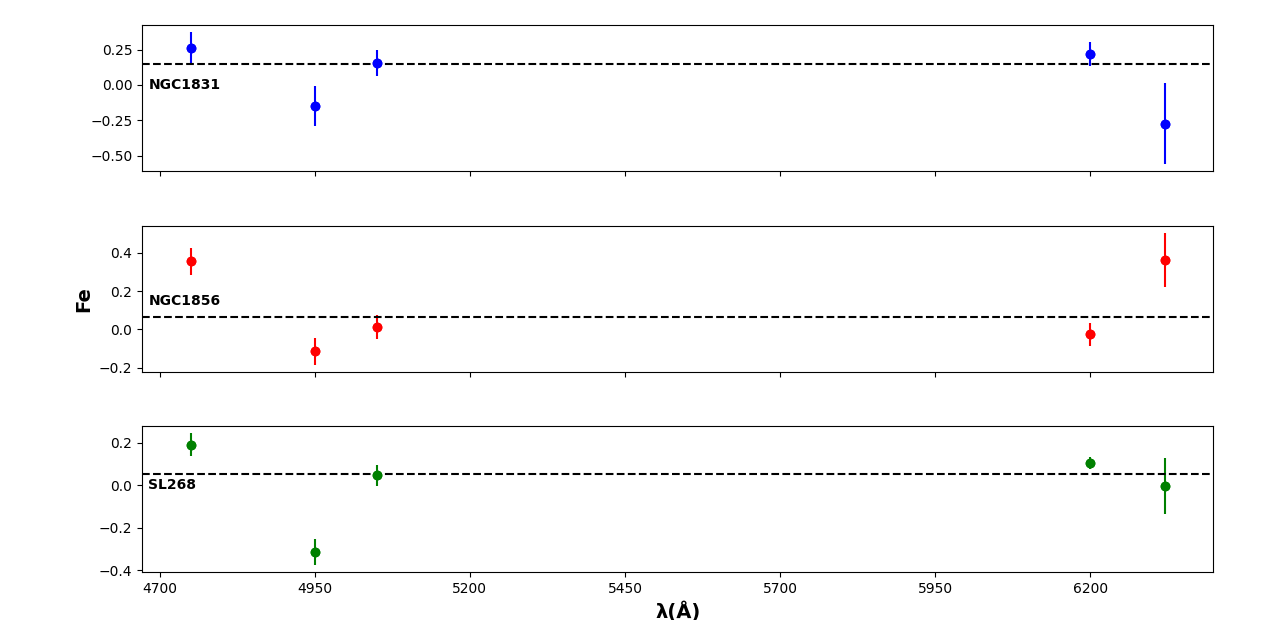}
\caption{Fe abundances listed in Tables \ref{Table_NGC1831}, \ref{Table_NGC1856} and \ref{Table_SL268} as a function of wavelength.The dotted line is the weighted average.}
\label{Fe}
\end{figure*}

\begin{figure*}
\plotone{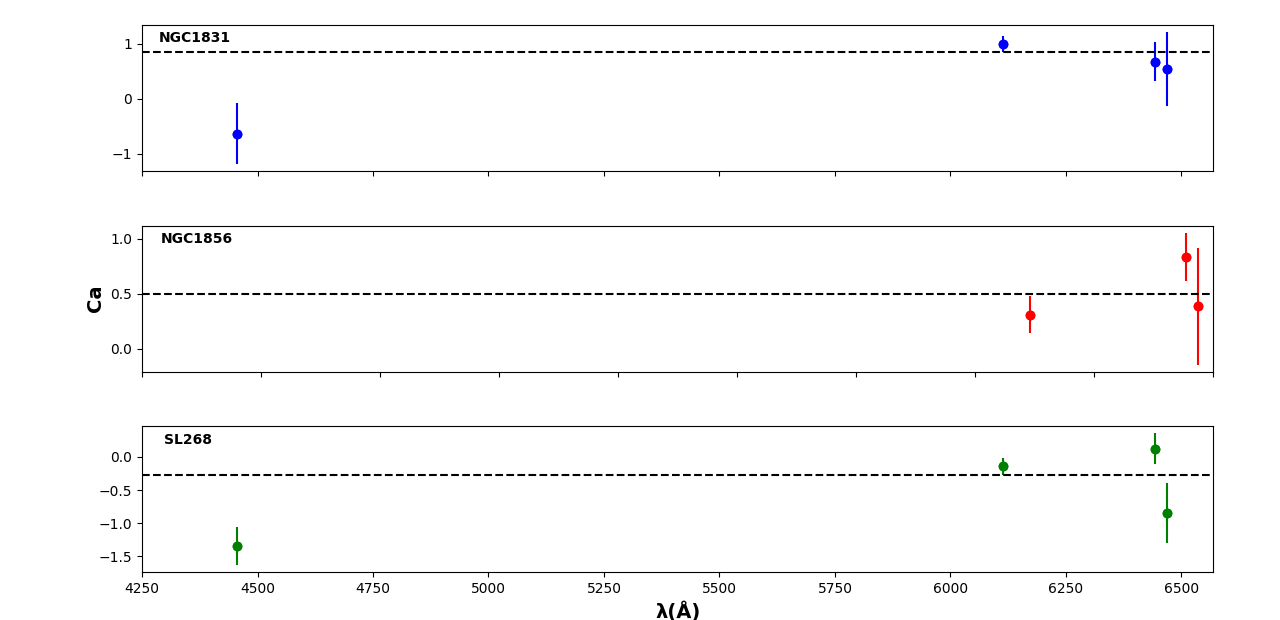}
\caption{Ca abundances listed in Tables \ref{Table_NGC1831}, \ref{Table_NGC1856} and \ref{Table_SL268} as a function of wavelength.The dotted line is the weighted average.}
\label{Ca}
\end{figure*}

\begin{figure*}
\plotone{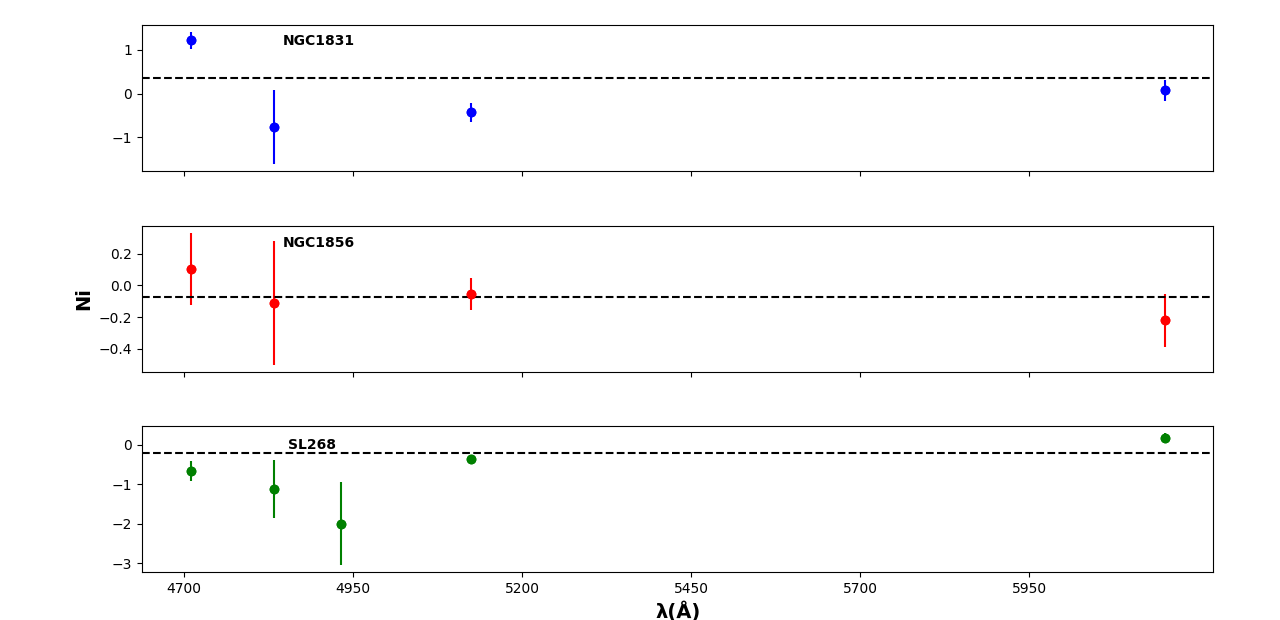}
\caption{Ni abundances listed in Tables \ref{Table_NGC1831}, \ref{Table_NGC1856} and \ref{Table_SL268} as a function of wavelength. The dotted line is the weighted average.}
\label{Ni}
\end{figure*}

\begin{figure*}
\plotone{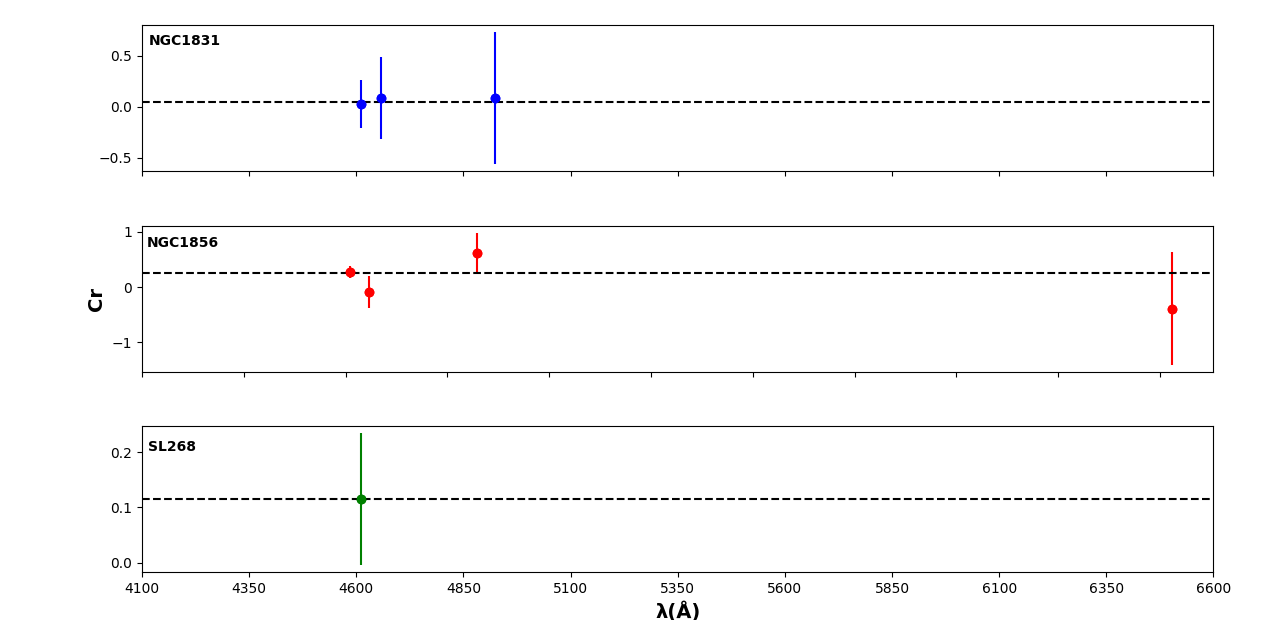}
\caption{Cr abundances listed in Tables \ref{Table_NGC1831}, \ref{Table_NGC1856} and \ref{Table_SL268} as a function of wavelength.The dotted line is the weighted average.}
\label{Cr}
\end{figure*}

In Tables \ref{Table_NGC1831}-\ref{Table_SL268} we list the individual bin measurements for the YMCs studied here. To aid in the assessment of the degree of dispersion between the different bins for a given element, in Figures \ref{Fe}-\ref{Cr} we plot the inferred abundances as a function of wavelength. 

\begin{table}
\caption{Chemical Abundances for NGC1831}
\label{Table_NGC1831}
\begin{tabular}{ccc}
\hline
Wavelength & Abundance & Error \\
\hline
[Z] \\
4000.00 - 4200.00 & -0.173 & 0.102 \\
4200.00 - 4400.00 & -0.215 & 0.058 \\
4400.00 - 4600.00 & -0.447 & 0.070 \\
4600.00 - 4800.00 & -0.257 & 0.104 \\
4800.00 - 5000.00 & -0.680 & 0.085 \\
5000.00 - 5200.00 & -0.550 & 0.053 \\
5200.00 - 5400.00 & -0.605 & 0.085 \\
5400.00 - 5550.00 & -0.188 & 0.104 \\
5600.00 - 5800.00 & -0.646 & 0.084 \\
Fe (dex)\\
4700.00 - 4800.00 & +0.265 & 0.110 \\
4900.00 - 5000.00 & -0.146 & 0.142 \\
5000.00 - 5100.00 & +0.154 & 0.091 \\
6100.00 - 6300.00 & +0.217 & 0.086 \\
6300.00 - 6340.00 & -0.274 & 0.289 \\
Ca (dex)\\
4445.00 - 4465.00 & -0.633 & 0.559 \\
6100.00 - 6128.00 & +0.995 & 0.141 \\
6430.00 - 6454.00 & +0.676 & 0.354 \\
6459.00 - 6478.00 & +0.546 & 0.668 \\
Cr (dex)\\
4580.00 - 4640.00 & +0.026 & 0.239 \\
4640.00 - 4675.00 & +0.085 & 0.400 \\
4915.00 - 4930.00 & +0.084 & 0.651 \\
6600.00 - 6660.00 & - & - \\
Mg (dex)\\
5150.00 - 5200.00 & +0.125 & 0.070 \\
Mn (dex)\\
4750.00 - 4770.00 & +0.366 & 0.350 \\
Na (dex)\\
5670.00 - 5700.00 & -0.531 & 0.536 \\
6148.00 - 6168.00 & +0.565 & 0.490 \\
Ni (dex)\\
4700.00 - 4720.00 & +1.215 & 0.191 \\
4825.00 - 4840.00 & -0.765 & 0.851 \\
4910.00 - 4955.00 & - & - \\
5075.00 - 5175.00 & -0.422 & 0.218 \\
6100.00 - 6200.00 & +0.074 & 0.231 \\
Ti (dex)\\
4650.0 - 4718.00 & -0.265 & 0.421 \\
4980.0 - 5045.00 & +0.635 & 0.216 \\
6584.0 - 6780.00 & +0.865 & 0.267 \\
\hline
\end{tabular}
\end{table}

\begin{table}
\caption{Chemical Abundances for NGC1856}
\label{Table_NGC1856}
\begin{tabular}{ccc}
\hline
 Wavelength & Abundance & Error \\
\hline
[Z] \\
4000.00 - 4200.00 & -0.381 & 0.062 \\
4200.00 - 4400.00 & -0.732 & 0.058 \\
4400.00 - 4600.00 & -0.792 & 0.066 \\
4600.00 - 4800.00 & -0.388 & 0.061 \\
4800.00 - 5000.00 & -0.543 & 0.048 \\
5000.00 - 5200.00 & -0.723 & 0.042 \\
5200.00 - 5400.00 & -0.700 & 0.050 \\
5400.00 - 5550.00 & -0.487 & 0.062 \\
5600.00 - 5800.00 & -0.421 & 0.051 \\
Fe (dex)\\
4700.00 - 4800.00 & +0.356 & 0.071 \\
4900.00 - 5000.00 & -0.115 & 0.070 \\
5000.00 - 5100.00 & +0.012 & 0.063 \\
6100.00 - 6300.00 & -0.024 & 0.060 \\
6300.00 - 6340.00 & +0.365 & 0.141 \\
Ca (dex)\\
4445.00 - 4465.00 & - & - \\
6100.00 - 6128.00 & +0.307 & 0.168 \\
6430.00 - 6454.00 & +0.835 & 0.221 \\
6459.00 - 6478.00 & +0.385 & 0.533 \\
Cr (dex)\\
4580.00 - 4640.00 & +0.275 & 0.110 \\
4640.00 - 4675.00 & -0.086 & 0.291 \\
4915.00 - 4930.00 & +0.615 & 0.371 \\
6600.00 - 6660.00 & -0.388 & 1.017 \\
Mg (dex)\\
5150.00 - 5200.00 & +0.045 & 0.031 \\
Mn (dex)\\
4750.00 - 4770.00 & -0.144 & 0.294 \\
Na (dex)\\
5670.00 - 5700.00 & -0.072 & 0.287 \\
6148.00 - 6168.00 & +0.465 & 0.271 \\
Ni (dex)\\
4700.00 - 4720.00 & +0.105 & 0.230 \\
4825.00 - 4840.00 & -0.111 & 0.392 \\
4910.00 - 4955.00 & - & - \\
5075.00 - 5175.00 & -0.055 & 0.101 \\
6100.00 - 6200.00 & -0.221 & 0.167 \\
Ti (dex)\\
4650.00 - 4718.00 & -0.004 & 0.209 \\
4980.00 - 5045.00 & +0.515 & 0.136 \\
6584.00 - 6780.00 & +0.428 & 0.177 \\
\hline
\end{tabular}
\end{table}
\begin{table}
\caption{Chemical Abundances for [SL63]268}
\label{Table_SL268}
\begin{tabular}{ccc}
\hline
 Wavelength & Abundance & Error \\
\hline
[Z] \\
4000.00 - 4200.00 & -0.326 & 0.025 \\
4200.00 - 4400.00 & -0.453 & 0.032 \\
4400.00 - 4600.00 & -0.712 & 0.041 \\
4600.00 - 4800.00 & -0.534 & 0.047 \\
4800.00 - 5000.00 & -0.459 & 0.039 \\
5000.00 - 5200.00 & -0.512 & 0.027 \\
5200.00 - 5400.00 & -0.570 & 0.038 \\
5400.00 - 5550.00 & -0.561 & 0.054 \\
5600.00 - 5800.00 & -0.464 & 0.043 \\
Fe (dex)\\
4700.00 - 4800.00 & +0.191 & 0.055 \\
4900.00 - 5000.00 & -0.316 & 0.062 \\
5000.00 - 5100.00 & +0.046 & 0.050 \\
6100.00 - 6300.00 & +0.105 & 0.030 \\
6300.00 - 6340.00 & -0.005 & 0.131 \\
Ca (dex)\\
4445.00 - 4465.00 & -1.346 & 0.291 \\
6100.00 - 6128.00 & -0.144 & 0.125 \\
6430.00 - 6454.00 & +0.126 & 0.235 \\
6459.00 - 6478.00 & -0.846 & 0.451 \\
Cr (dex)\\
4580.00 - 4640.00 & +0.115 & 0.120 \\
4640.00 - 4675.00 & - & - \\
4915.00 - 4930.00 & - & - \\
6600.00 - 6660.00 & - & - \\
Mg (dex)\\
5150.00 - 5200.00 & +0.074 & 0.031 \\
Mn (dex)\\
4750.00 - 4770.00 & -0.195 & 0.181 \\
Na (dex)\\
5670.00 - 5700.00 & +0.316 & 0.159 \\
6148.00 - 6168.00 & +0.433 & 0.202 \\
Ni (dex)\\
4700.00 - 4720.00 & -0.669 & 0.244 \\
4825.00 - 4840.00 & -1.126 & 0.731 \\
4910.00 - 4955.00 & -1.997 & 1.057 \\
5075.00 - 5175.00 & -0.356 & 0.092 \\
6100.00 - 6200.00 & +0.174 & 0.121 \\
Ti (dex)\\
4650.00 - 4718.00 & +0.081 & 0.185 \\
4980.00 - 5045.00 & +0.354 & 0.091 \\
6584.00 - 6780.00 & -0.188 & 0.211 \\
\hline
\end{tabular}
\end{table}

\end{document}